\documentclass[10pt,letterpaper,twocolumn]{article} 

\usepackage{ol2}
\usepackage[draft,implicit=false]{hyperref}
\usepackage{amsmath}

\begin{document}

\twocolumn[ 

\title{Barium Fluoride Whispering-Gallery Mode Disk-Resonator \\
       with one Billion Quality-Factor}
\author{Guoping Lin$^*$, Souleymane Diallo, R\'emi Henriet, Maxime Jacquot,\\
        and Yanne K. Chembo}
\address{FEMTO-ST Institute [CNRS UMR6174], Optics Department, \\
         15B Avenue des Montboucons, 25030 Besan\c con cedex, France.\\
         $^*$Corresponding author: {guoping.lin@femto-st.fr}
         }

\begin{abstract}
We demonstrate a monolithic optical whispering gallery mode resonator fabricated with barium fluoride (BaF$_2$) with an ultra-high quality ($Q$) factor above $10^9$ at $1550$~nm, and measured with both the linewidth and cavity-ring-down methods. 
Vertical scanning  optical profilometry shows that the root mean square surface roughness of $2$~nm is achieved for our mm-size disk.
To the best of our knowledge, we show for the first time that one billion $Q$-factor is achievable by precision polishing in relatively soft crystals with mohs hardness of~$3$.  
We show that complex thermo-optical dynamics can take place in these resonators.
Beside usual applications in nonlinear optics and microwave photonics,
high energy particle scintillation detection utilizing monolithic BaF$_2$ resonators potentially becomes feasible.
\end{abstract}
]


Monolithic whispering gallery mode (WGM) resonators are idoneous platforms to study various properties of optical materials. These resonators have attracted large interest, as they can feature ultra-high $Q$ factors and small mode volumes~\cite{vahala2003optical}. 
Light in such resonators is trapped by a process of successive total internal reflections at the inner side of the disk circumference.
Therefore, they do not require additional precision optical coating and are able to reach $Q$ factors close to the
material absorption limit. 

Beside the surface tension~\cite{vahala2003optical} and chemical etching~\cite{lee2012chemically} techniques that have been reported to fabricate ultra-high $Q$ resonators (above 100~million at $1550$~nm), mechanical polishing methods can be used for various host materials, especially optical crystals~\cite{savchenkov2004kilohertz}. Owing  to the low absorption of crystalline materials, extremely high $Q$ (in the order of $10^{11}$) have already been demonstrated with calcium fluoride~\cite{savchenkov2007optical}. Many crystalline materials such as quartz~\cite{ilchenko2008crystal}, sapphire~\cite{ilchenko2014Sapphire}, calcium fluoride~\cite{grudinin2006ultra,Savchenkov093902PRL,Grudinin:07,grudinin2009brillouin, grudinin2009generation}, magnesium fluoride~\cite{tavernier2010magnesium,Liang11,herr2014temporal}, lithium niobate~\cite{ilchenko2004nonlinear,furst2010naturally}, beta barium borate~\cite{lin2012high,lin2013wide} and more recently diamond~\cite{Ilchenko:13} have also been used to manufacture ultra-high $Q$ WGM resonators. 
These resonators can be used for various applications such as Raman lasing~\cite{Grudinin:07}, Brillouin lasing~\cite{grudinin2009brillouin}, second-harmonic generations~\cite{ilchenko2004nonlinear,furst2010naturally,lin2013wide}, electro-optic modulators and microwave receivers~\cite{ilchenko2003whispering}, frequency comb generation~\cite{Savchenkov093902PRL, grudinin2009generation, Liang11,herr2014temporal}, optoelectronic microwave oscillators~\cite{volyanskiy2010compact, maleki2011sources}, laser locking and stabilization ~\cite{liang2010whispering}, amongst others.      

Barium fluoride is a particularly interesting optical material which is transparent from ultraviolet to mid-infrared. It is the most resistant to high energy radiation among all fluoride crystals. 
It is also known as one of the fastest scintillators for the detection of high energy particles such as X-rays and gamma rays~\cite{laval1983barium}. However, to this date, few efforts have been made to study this crystal using optical resonator structures.
So far, $Q$ factors above $10^9$ at $1550$~nm have only been demonstrated on resonators made from four optical crystals, namely
quartz, calcium fluoride (CaF$_2$), magnesium fluoride (MgF$_2$) and sapphire. It should be noted that these crystals have mohs scale material hardness between $4$ and $9$. It still remains unknown whether the same order of $Q$ factors can be obtained on softer crystals like BaF$_2$ whose mohs hardness is only~$3$, a feature that is of uttermost importance at the time to consider large-scale automated fabrication. 

\begin{figure}[t]
	\centering
	\includegraphics[width=7.5cm]{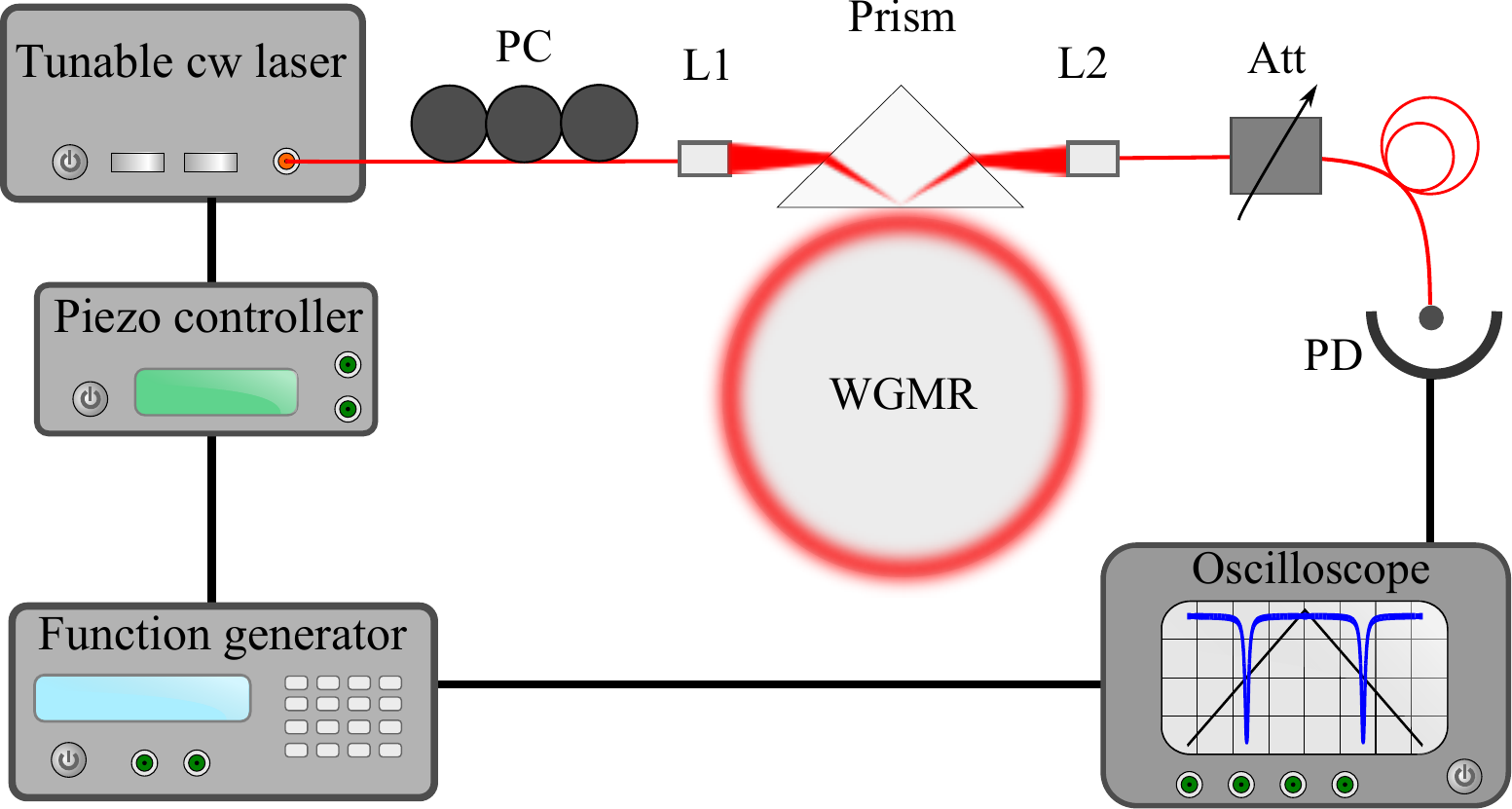}	
	\caption{ (Color online) Schematic illustration of the experimental setup for the WGM resonator characterization.
	PC: fiber polarization controller; 
	L1, L2: GRIN lenses; 
	Att: attenuator;
	PD: photodetector.}
	\label{fig:setup}
\end{figure}

In this letter, we show for the first time that one billion $Q$ factor is achievable in optical resonators made from relatively soft crystal with mohs hardness of~$3$ using precision polishing. 
The experimental setup presented in Fig.~\ref{fig:setup} has been used to determine accurately the value of this 
quality factor. 
We have also performed a vertical scanning profilometry on the rim of the polished disk, that enabled us to demonstrate a root mean square surface roughness of $2$~nm. 
We also report the observation of different thermal distortions in such crystal, which originate from the interplay between 
the negative thermo-optic coefficient and other positive effects such as thermal expansion and Kerr nonlinearity. 

\begin{figure}[b]
	\centering
	\includegraphics[width=7.5cm]{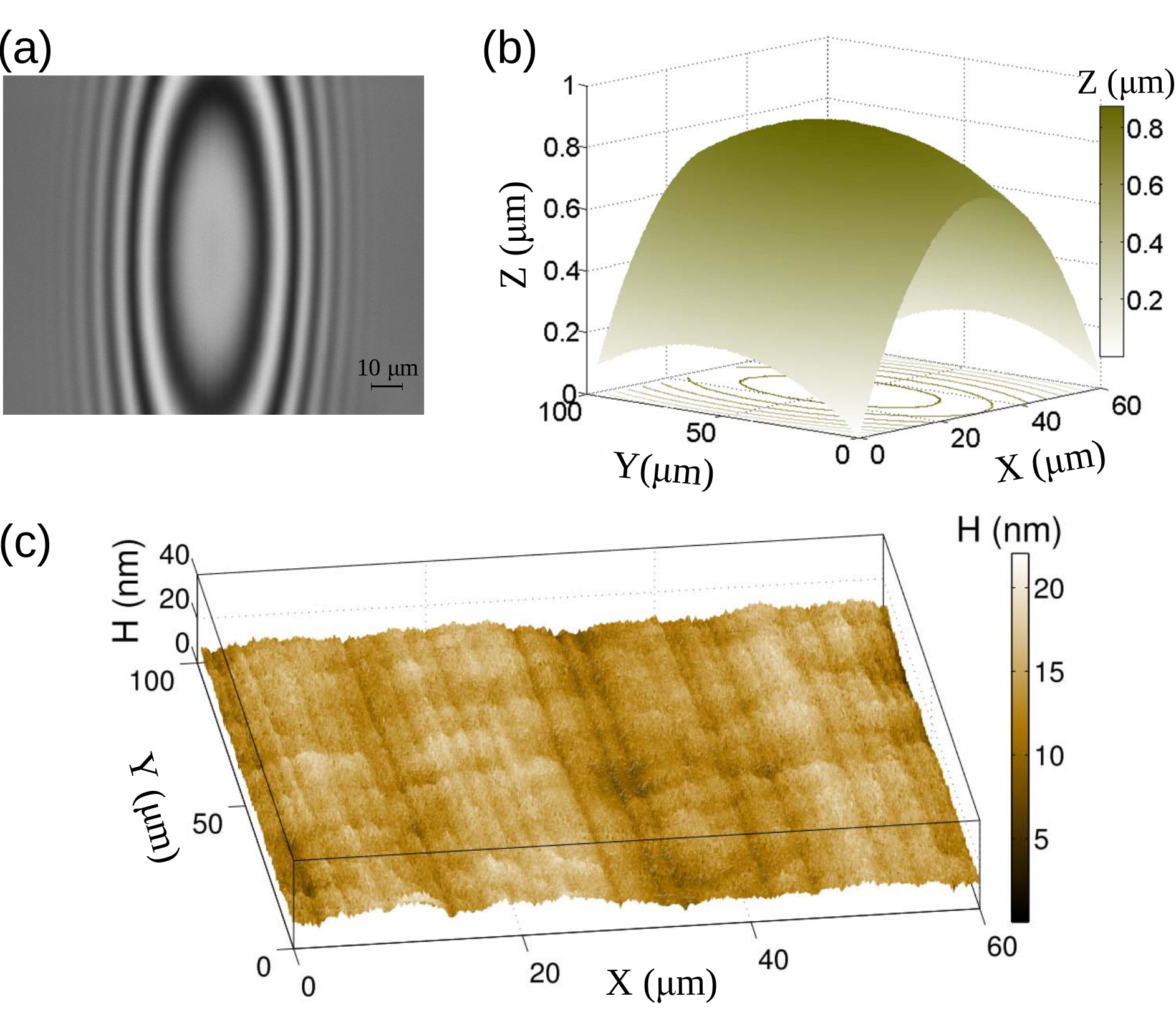}
	
	\caption{ (Color online) {(a)}~An example of interference pattern from the edge surface on an ultra-high $Q$ BaF$_2$ disk, taken with a vertical scanning microscope at $40 \times$ magnification. 
	         {(b)}~Reconstructed 3D~surface profile. 
	         {(c)}~Reconstructed surface roughness profile, showing rms surface roughness of $2$ nm. }
	\label{fig:profilometry}
\end{figure}

We have fabricated the ultra-high $Q$ WGM resonators from commercially available BaF$_2$ disks with diameters of $12$~mm and thicknesses of $1$~mm. 
To monitor the surface profile of BaF$_2$ resonators, we carried out surface profile measurements using white light vertical scanning interferometry with conventional phase shifting algorithm. 
Figure ~\ref{fig:profilometry}(a) presents a white light interference pattern on the edge of the disk at $40\times$~magnification with a Mirau objective. It should be mentioned that a spherical surface produces circular symmetric fringes. The elliptic fringes comes from a toroidal or oblate surface. 
An example of the computer reconstructed 3D~surface profile on the resonator with $11.6$ mm diameter is presented in 
Fig.~\ref{fig:profilometry}(b). It covers a surface area of $100~\mu$m $\times 60~\mu$m. 
We can thereby determine the rms surface roughness profile as shown in Fig.~\ref{fig:profilometry}(c). 
The measured rms surface roughness is about $2$ nm ($\sim \lambda/800$ at the telecom wavelength window). 
Straight nano-sctrach lines created during the polishing process are also observed (depth $\sim 10$~nm).
It should be noted that the fundamental (torus-like) mode in a mm-size disk resonator usually features a spot size less than
$20~\mu$m in its cross section, and as a result, these lines could cause additional scattering losses. 
On the other hand, if they are created in the direction parallel to the symmetry axis of the disk, grating effects could also be observed~\cite{aveline2013whispering}.

After confirming that we have achieved a good surface roughness, we move forward to WGM excitation experiments. 
The refractive index of BaF$_2$ at $1550$ nm is $1.466$ which is slightly larger than that of a regular optical fiber. 
As a result, fiber tapers can not be used to efficiently excite high $Q$ WGMs due to phase mismatch, although it is a convenient way to quickly probe the resonator without the risk of scratching its surface. 
Nevertheless, one can choose another evanescent wave coupling method by using an optical prism that has a higher refractive index. In this experiment, a SF11 prism is used.
 
Figure~\ref{fig:setup} shows the schematic illustration of our experimental setup. The excitation source is a tunable continuous-wave (cw) fiber laser at $1550$ nm with sub-kHz instantaneous linewidth. The laser frequency is scanned using a piezo driver controlled by the function generator. A fiber polarization controller is added before launching the light into the prism. The gradient-index (GRIN) lens (L1) focuses the laser beam on the prism, while the other one collects the output beam back into a single mode fiber for detection. A piezoelectric actuator is inserted into a translation stage to finely control the coupling gap between the resonator and the prism.

There are two methods to evaluate the $Q$ factors of a resonator, which are namely linewidth and ring-down measurements. 
In this letter, we carry out both measurements on our BaF$_2$ resonators. 

\begin{figure}[tp]
	\centering
	\includegraphics[width=7.5cm]{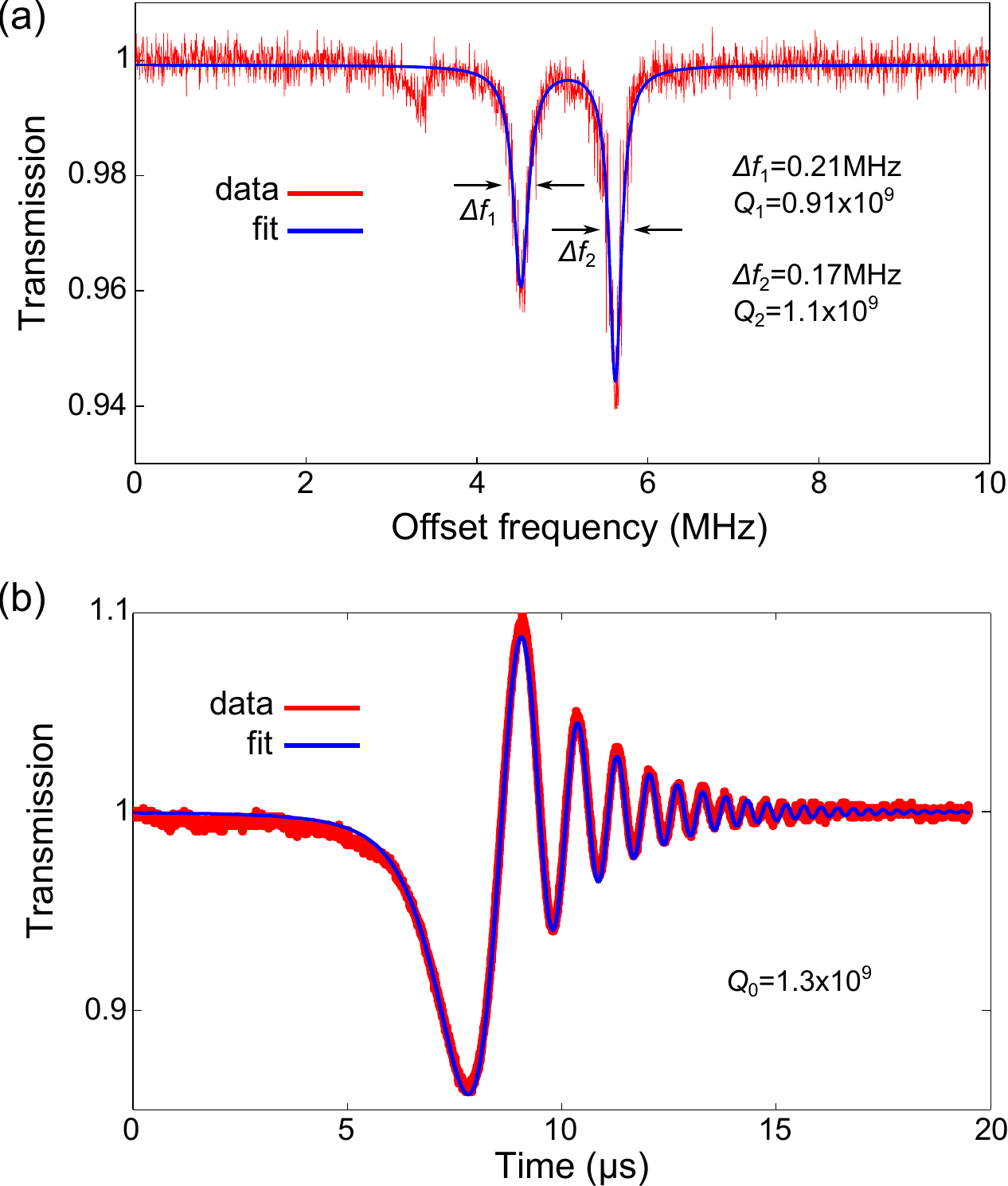}
	\caption{ (Color online) {(a)}~Transmission spectrum of WGM resonances. 
	               The Lorentzian fit yields a linewidth $\Delta f$ as narrow as $0.17$ MHz, corresponding to a $Q$ factor
	               of $1.1 \times 10^9$. 
	        {(b)} Transmission spectrum with a ring-down structure.
	               A theoretical fit provides an intrinsic $Q$ factor of $1.3 \times10^9$.}
	\label{fig:Q}
\end{figure}

Figure~\ref{fig:Q}(a) shows a typical transmission spectrum featured with the observed ultra-high $Q$ WGMs from the oscilloscope. The laser frequency is scanned across the optical resonances at the speed of $1.1$~GHz/s and with an incident power of $9.5$~mW. The coupling gap is set to be large enough so that the resonances are strongly under-coupled.
In this case, the coupling losses are negligible and the measured resonance linewidth approaches the intrinsic one.
We also make sure that we do not see visible thermal distortions at both up and down detuning directions. 
The Lorentzian fit on the data is shown in Fig.~\ref{fig:Q}(a). A linewidth $\Delta f$ as narrow as $0.17$ MHz is obtained. The corresponding $Q$ factor is  $1.1 \times 10^9$ using the formula $Q=f_0/\Delta f$. 
It should be mentioned that the double resonances with splitting of $1.1$ MHz could result from stimulated Rayleigh scattering in the resonator which is often observed in ultra-high $Q$ resonators~\cite{Weiss:95,Gorodetsky2000Rayleigh}. 

The ring-down spectroscopy can be performed by switching off the probe laser using a fast shutter. It is also known that ring-down can be observed when the laser frequency is swept across the resonance at a fast speed~\cite{savchenkov2007optical,trebaol2010ringing}. 
The light accumulated in the mode is re-emitted and interferes with the laser through-put in the exit port. 
In order to observe this phenomenology, we increase the modulation rate and frequency range of the laser. 
With the detuning speed set at $65$~GHz/s, a WGM ringdown spectrum is obtained as shown in Fig.~\ref{fig:Q}(b).
The theoretical fit gives an intrinsic  $Q$ value of $1.3 \times 10^9$. 
We have also excited  WGM families with an orthogonal polarization and observed a smaller $Q$ factor of $1.1 \times 10^9$.
The $Q$ factors differ by a factor of $1.2$, which is smaller than $n_0^4$. Therefore, we believe that the cleaning of the resonator is sufficient for the $Q$ factors observed~\cite{Ilchenko:13}.

The $Q$~factor measurement is also an interesting method to evaluate the material absorption coefficient in highly transparent crystals.  Theoretically, the intrinsic factor $Q_0$ of a resonator is mainly determined by three different loss mechanisms. The corresponding limited $Q$ factors are surface scattering loss related $Q_{ss}$, material absorption loss related $Q_\alpha$ and radiation loss related $Q_r$. The final $Q_0$ is then determined by $1/Q_0=1/Q_{ss}+1/Q_\alpha+1/Q_r$. In the case of a mm-size resonator with a good circular symmetry, the last term $1/Q_r$  is usually negligible. Concerning surface scattering limited $Q_{ss}$, under the condition that the surface roughness is much smaller than the optical wavelength, this value can be estimated using~\cite{Gorodetsky2000Rayleigh}:
\begin{equation}
  Q_{ss} \approx \frac{3\lambda^3 R}{8 \pi^2 n_0 B^2 \sigma^2}
    \label{eq:SSQ}
\end{equation}
where $\lambda$ is the wavelength, $R$ is the radius of the resonator, $n_0$ is the refractive index of the material, $\sigma$ and $B$ are the surface roughness and the correlation length of the roughness. Assuming that $B$ is equal to $\sigma=2$ nm, we obtain an estimated $Q_{ss}$ value of $3.7\times 10^{13}$ at the wavelength of $1550$~nm for a radius of $6$ mm. This value is far larger than the experimental one of $1.3 \times 10^9$ as shown in Fig.~\ref{fig:Q}. Even we use $B=\sigma=10$ nm, the estimated $Q_{ss}$ of $5.8 \times 10^{10}$ is still much larger. Although we didn't check the whole periphery of the mm size resonator which is a lengthy task, we expect that no deep scratches appear on the surface and the material absorption is the dominant factor that limits $Q_0$.

On the other hand, the material absorption limited $Q$ is expressed as:
\begin{equation}
  Q_{\alpha}=2 \pi n_0/(\lambda a)
    \label{eq:absorpQ}
\end{equation}
where $n_0=1.466$ is the refractive index of BaF$_2$ at $1550$~nm. Therefore, we can derive the material absorption coefficient of $4.6\times10^{-5}$ cm$^{-1}$ at this wavelength.

We also report interesting thermal dynamic processes observed in BaF$_2$ resonators.
In WGM resonators, the resonance frequency shift $\Delta f(t)$ obeys the following formulation:
\begin{equation}
  \frac{\Delta f(t)}{f_0}=- \left[\frac{1}{n_0} \frac{dn}{dT}\Delta T_1+\frac{1}{R}\frac{dR}{dT} \Delta T_2(t)
                                +\frac{n_2}{n_0}\frac{P_c(t)}{A_{\rm eff}} \right]
    \label{eq:Rshift}
\end{equation}
where $f_0$ is the cold cavity resonance frequency, $n_2$ is the Kerr coefficient, $\Delta T_1$ is the temperature change in the mode area, $\Delta T_2$ is the temperature change in the whole resonator, $dn/dT$ and $(dR/R)/dT$ represent the thermo-optic ($-16 \times 10^{-6}/K$) and thermal expansion ($18.4 \times 10^{-6}/K$) coefficients  of BaF$_2$~\cite{weber2002handbook}, $P_c(t)$ designates the intracavity optical power, $A_{\rm eff}$ is the cross section mode area. It should be noted that the second term on the right side is a very slow  time-scale phenomenon, as the optical heating requires time to diffuse heat into the full  volume in the case of a mm-size resonator.
Nevertheless, it is interesting to note that the interplay of positive and negative coefficients together with slow and fast phenomena could lead to complex thermo-optical dynamics~\cite{Deng13,he2009oscillatory}. 

\begin{figure}[t]
	\centering
	\includegraphics[width=7.5cm]{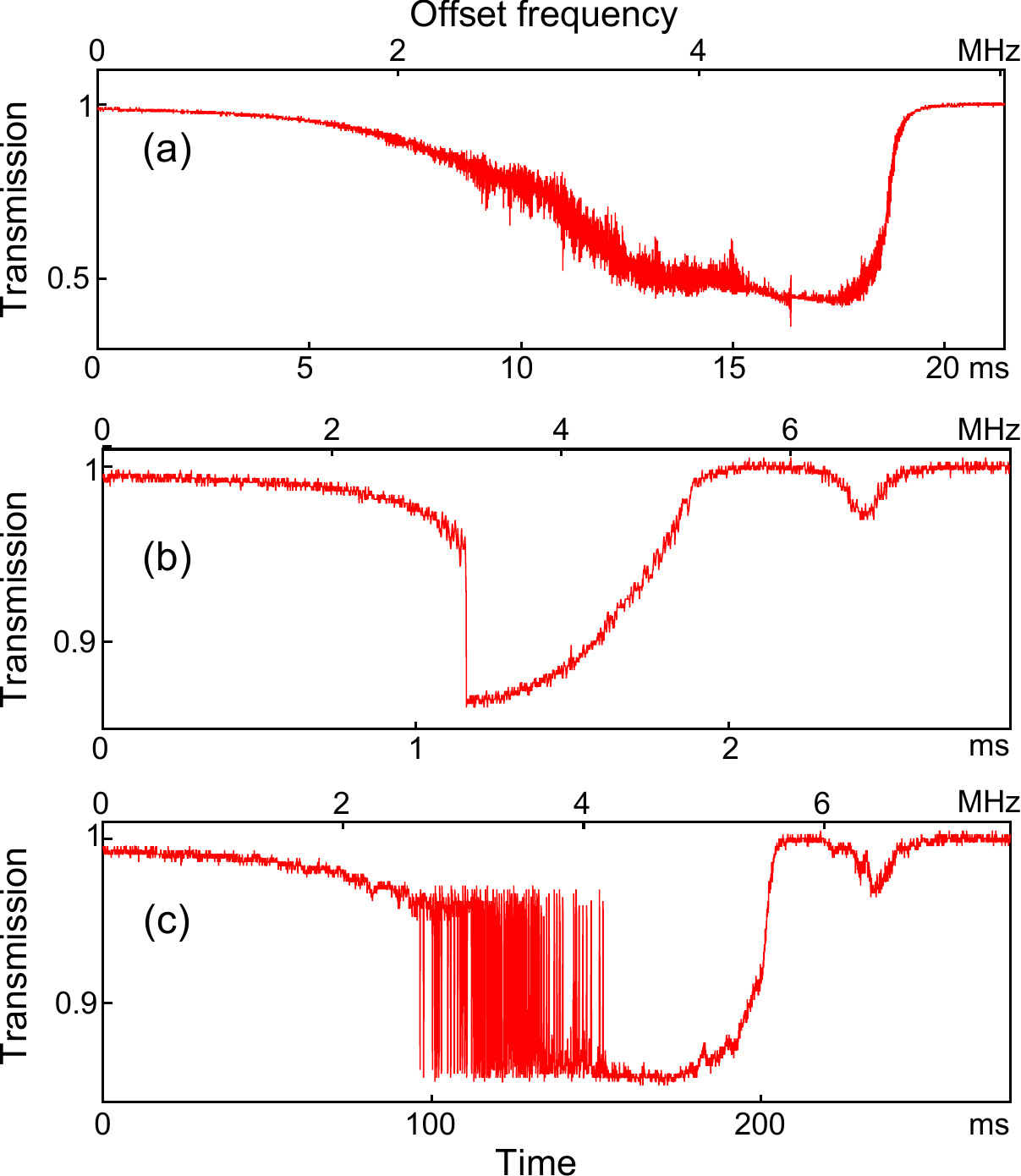}
	
	\caption{ (Color online) Comparison of different thermal distortions in BaF$_2$ resonators.
	        (a) Incident pump power: $18$~mW, detuning speed: $0.27$~GHz/s;
	        (b) Incident pump power: $45$~mW, detuning speed: $2.7$~GHz/s;
	        (c) Incident pump power: $45$~mW, detuning speed: $27$~MHz/s.}
	\label{fig:thermal}
\end{figure}

Figure~\ref{fig:thermal}(a) shows a typical thermal distortion in a crystalline resonator dominated by its negative thermo-optic coefficient. The increasing of the laser frequency (decreasing the wavelength) is followed by the simultaneous blue shift of the resonance, which is interpreted by the first term on the right side of Eq.~(\ref{eq:Rshift}). 
As a result, the resonance is broadened on the short frequency side of the center resonance when the laser frequency detuning is increasing. This effect has been used for fast microlaser characterizations~\cite{lin2012thermal}. 
We also observed an interesting thermal distortion in a different optical mode as shown in Fig.~\ref{fig:thermal}(b), where the larger frequency side is broadened. The laser was swept at $2.7$~GHz/s, that is, $10$~times faster. 
Hence, we believe that Kerr nonlinearity is dominant in this case. 
We also observed oscillatory behaviors on the same mode when we further reduced the ramp speed to $27$~MHz/s.
In this case, the interplay of positive and negative coefficients causes an oscillatory behavior, as shown in Fig.~\ref{fig:thermal}(c).
Similar oscillation phenomenon was reported in hybrid resonators aiming for self-thermo compensations~\cite{he2009oscillatory}.


In conclusion, we have reported for the first time a  monolithic BaF$_2$ WGM resonator with one billion $Q$ factor. 
It is the softest material ever reported featuring with such $Q$ factors. We also derive a new upper bound of material absorption coefficient at $1550$ nm. Interesting thermal effects are also observed. These observation will benefit future applications on stable frequency comb generation, Brillouin lasing and corresponding pure microwave and pulse laser generations. Considering the unique application of BaF$_2$ as a scintillator~\cite{laval1983barium} compared with other fluoride crystals, potential application of these monolithic resonators in high energy particle detection also becomes within reach.


The authors acknowledge financial support from the European Research Council (ERC) through the projects NextPhase and Versyt.
They also acknowledge financial support from the \textit{Centre National d'Etudes Spatiales} (CNES) through the project SHYRO, from the R\'egion de Franche-Comt\'e.  This project has been performed in cooperation with the Labex ACTION program.


\pagebreak

\section*{Informational Fourth Page}

\end{document}